\documentclass[aps,pra,floatfix,twocolumn,superscriptaddress,nofootinbib,a4paper]{revtex4}

\usepackage{graphicx}
\usepackage{graphics}
\usepackage{amssymb}
\usepackage{amsmath}
\usepackage{epsfig}
\usepackage{latexsym}
\usepackage{color}
\usepackage{subfigure}
\usepackage{array}

\newcommand{\nop}[1]{{}}
\newcommand{\eqn}[1]{\label{#1}}
\newcommand{\eq}[1]{Eq.~(\ref{#1})}
\newcommand{\cnot}{\textsc{cnot}}

\begin{document}

\title{Universal quantum dynamics from two 2-local Hamiltonians}

\author{Charles D. Hill}
\email{Charles.Hill@liverpool.ac.uk} \affiliation{Department of
Electrical Engineering and Electronics, University of Liverpool,
Brownlow Hill, Liverpool, L69 3GJ, United Kingdom}

\author{Henry L. Haselgrove}
\affiliation{C3I Division, Defence Science and Technology Organisation,
Edinburgh 5111 Australia}


\begin{abstract}
In this paper, we show that the ability to switch globally between two 2-local Hamiltonians on $n$ qubits is sufficient for achieving universal unitary dynamics on those $n$ qubits.  Of the two Hamiltonians used in the construction, one is comprised of nearest-neighbour two-qubit interactions on a one-dimensional chain, the other is comprised of single-qubit interactions. We describe a scheme for choosing the switching times in order to generate arbitrary one- and two-qubit gates on the $n$ qubits. However the switching scheme is inefficient, and we leave as an open problem the question of whether the universality of the Hamiltonians can be exploited in an efficient way.
\end{abstract}

\maketitle


\section{Introduction}

A fundamental question in quantum information science is what type of control on a many-qubit system is needed in order to efficiently and reliably achieve universal quantum dynamics.  Across the wide range of proposals for physically realizing a quantum computer, many different control schemes have been envisaged (a handful of such include \cite{KLM01,NPT98,CZ95,LD98,IAB+99,GC97,CFH97,Kan98} ).
A typical requirement is that one- and two-qubit interactions be controlled dynamically and independently on the various qubits in the system, thus directly enabling a universal set of one- and two-qubit gates on any subset of the qubits.

By comparison, the situation we are concerned with limits the available control of an $n$-qubit system to just two degrees of freedom. We imagine that the total system Hamiltonian $H_{tot}$ is comprised of two $n$-qubit terms $H_1$ and $H_2$ which remain constant over time, apart from the overall strength of each which can be varied over time and independently to the other:
\begin{equation}
H_{tot}(t) = h_1(t)H_1 + h_2(t)H_2. \eqn{htot}
\end{equation}
(We will only consider control functions $h_1(t)$ and $h_2(t)$ which are nonnegative-valued).
We show that this system is sufficient to generate universal unitary dynamics on the $n$ qubits, even when $H_1$ and $H_2$ have a form that is physically plausible: $H_1$ consists of single-qubit interactions, and $H_2$ consists of nearest-neighbour two-qubit interactions along a line.

A control scheme conforming to \eq{htot} has the appealing feature that
the functions $h_1(t)$ and $h_2(t)$ can always be transformed in a way that leaves the first one constant, yet such that the resulting unitary dynamics are the same as the original (or approximately the same, in the case where $h_1(t)$ has zeros). That is, a system having a ``constant + drift'' Hamiltonian
\begin{equation}
H'_{tot}(t) = H_1 + h'_2(t)H_2 \eqn{htot2}
\end{equation}
is essentially equivalent to a system governed by \eq{htot}. Understanding systems which achieve universal quantum dynamics via the control of a single global parameter can be valuable in theoretical applications\footnote{For further discussion and an example of such a theoretical use see for instance Section IV of \cite{Nie06}.}.
Practical applications may provide further motivation for understanding such systems. A control scheme of the type presented in this paper is one that could conceivably be implemented by varying a single field across a many-body system, and so in this sense may constitute less demanding physical requirements compared with traditional schemes that directly control each individual qubit.

There is a range of previous work which relates to generating universal quantum dynamics on many qubits when the control is limited to a few degrees of freedom. The results in \cite{Llo95, Wea00, BBC+95} show that almost any randomly selected pair of $n$-qubit
Hamiltonians $H_1$ and $H_2$ generates universal quantum dynamics. However, such randomly generated Hamiltonians will generally contain unphysical many-body terms; and it is also difficult to derive appropriate control functions. Lloyd {et al.} \cite{LLo04} showed that varying a single-qubit Hamiltonian term can in principle generate universal dynamics in a many-body system, although a method to construct control functions was not given. In \cite{Nie06} specific examples of Hamiltonians $H_1$ and $H_2$ are given that yield simple and efficient control functions for generating standard quantum gates ; although these Hamiltonians also contain unphysical many-body terms. Benjamin \cite{Ben01} has described an ingenious scheme for quantum computation which involves switching globally between {\em four} physically-realistic Hamiltonians. Control functions for that scheme are relatively simple and efficient. Fitzsimons and Twamley \cite{Fit06} have a simple and efficient scheme for quantum computing utilizing two physically-realistic global Hamiltonians plus the ability to individually address two particular qubits.  Quantum cellular automata \cite{Rau04} provide another approach to simple-control universal quantum computing,  where one step in the computation consists of a series of four-qubit unitaries applied in a translationally-invariant fashion on a grid. 

Our paper is organised as follows: Section \ref{sec:Universal}
defines the two 2-local Hamiltonians $H_1$ and $H_2$, and sketches a proof of their universality. Section \ref{sec:Method}
describes a constructive method for choosing the switching times between $H_1$ and $H_2$ in order to generate standard quantum gates. Conclusions are drawn in Section \ref{sec:Conclusion}.

\section{The two Hamiltonians} \label{sec:Universal}

In this section we define two $n$-qubit 2-local\footnote{A Hamiltonian is 2-local if it can be expressed as a sum of terms that each act on at most two bodies. Interactions which are not 2-local tend to be very difficult to construct in nature.} Hamiltonians $H_1$ and $H_2$
that are universal for quantum computation. That is, we show that for any $n$-qubit unitary $U$, we may achieve the evolution of $U$ to arbitrary accuracy by alternatively evolving the Hamiltonians $H_1$ and $H_2$ for some appropriate choice of times $t_1$, $t_2$, $t_3$, $\dots$, i.e.:
\begin{equation}
U \approx \exp(-i H_1 t_1) \exp(-i H_2 t_2) \exp(-i H_1 t_3)  \ldots   \eqn{universal}
\end{equation}
(Comparing with \eq{htot}, we are effectively limiting ourselves to control functions that obey $h_1(t)h_2(t)=0$ for all times $t$).


Universality has been considered for several years in the context of
quantum computing. To show that our choice of Hamiltonians are
universal, we will show that they can generate a universal gate set.
There are several well known universal gate sets; we
will generate the set consisting of the Hadamard, ``$\pi/8$'', and controlled-not gates (denoted $H$, $T$, and \cnot), where
\begin{eqnarray}
H &=& \textstyle\frac{X+Z}{\sqrt{2}}, \\
T &=& e^{-i\frac{\pi}{8}Z}, \\
\cnot &=& |00\rangle\langle 00| + |01\rangle\langle 01| \\&& + |10\rangle\langle 11| + |11 \rangle\langle 10|,
\end{eqnarray}
and where $X$ and $Z$ are the Pauli operators $X=|0\rangle\langle 1| + |1\rangle\langle 0|$ and $Z=|0\rangle\langle 0| - |1\rangle\langle1|$.

Consider the Hamiltonian
\begin{equation}
H_H = \sum_{m=1}^n \frac{X_m + Z_m}{\sqrt{2}}, \eqn{hamh}
\end{equation}
where $X_m$ and $Z_m$ denote the operators $X$ and $Z$ acting on the $m$-th qubit.
Evolving this Hamiltonian for $t=\pi/2$ (or more generally $t=\pi/2+k\pi$ for any  $k\in \mathbb{Z}$) will implement a Hadamard on
every qubit,
\[
U_H = H \otimes H \otimes H \otimes H \ldots,
\]
whereas evolving the Hamiltonian for $t=k\pi$, $k\in \mathbb{Z}$, will instead implement the identity on every qubit,
\[
U_I = I \otimes I \otimes I \otimes I \ldots
\]
Now, we require the ability to apply a
Hadamard selectively to a qubit of our choice.
To do this we alter \eq{hamh} so that the terms that act on each qubit are given different (but fixed) relative strengths to one another:
\begin{equation}
H_1 = \sum_{m=1}^n a_m \frac{X_m + Z_m}{\sqrt{2}}. \eqn{h1}
\end{equation}
The coefficients $a_m$ should be chosen to be incommensurate with
each other; that is no pair of $a_m$ should have a ratio that is a rational number. For example, we could choose $a_1=1$, $a_2=\sqrt{5}$,
$a_3 = e$, and so on.

This allows us to isolate a particular term. As $H_1$ evolves over time, the overall evolution on each qubit becomes equal to $I$ or $H$ alternatively at points in time separated by $\pi/2/a_m$. Since this period is different for each qubit, one simply has to wait long enough for a {\em coincidence} to occur, whereby the desired qubit has evolved by approximately $H$ and the other qubits have evolved by approximately $I$. That is, there exists times $t_1$, $t_2$, \dots such that
\begin{eqnarray}
U_{H_1}(t_1) &\approx& H \otimes I \otimes I \otimes I \ldots, \label{eqn:HII}\\
U_{H_1}(t_2) &\approx& I \otimes H \otimes I \otimes I \ldots \label{eqn:IHI} \\
 \cdots & & \mbox{and so on.}
\end{eqnarray}
The longer one is prepared to wait, the better the ``coincidences'' that will be found, and thus more accurate gates will result. In other words, choosing irrational
coefficients guarantees the the space of $n$ single-qubit rotations about the
$(X+Z)$-axis is densely filled by the trajectory of $e^{-iH_1t}$.


The second Hamiltonian $H_2$ is chosen as follows,
\begin{equation}
H_2 = \sum_{m=1}^n b_m Z_m + \sum_{m=1}^{n-1} c_m Z_m \otimes Z_{m+1}, \eqn{h2}
\end{equation}
where the coefficients $b_m$ and $c_m$ are incommensurate with one another. Note that like \eq{h1}, all terms in \eq{h2} commute with one another. The arguments used above can again be used to show that evolving $H_2$ for some appropriate times $t'_1$, $t'_2$, \dots,  achieves the approximate evolution of the gate $T=e^{-i\frac{\pi}{8}Z}$ on any qubit,
\begin{eqnarray}
U_{H_2}(t'_1) &\approx& e^{-i\frac{\pi}{8}Z} \otimes I \otimes I \otimes I \ldots, \label{eqn:TII}\\
U_{H_2}(t'_2) &\approx& I \otimes e^{-i\frac{\pi}{8}Z} \otimes I \otimes I \ldots \label{eqn:ITI} \\
 \vdots & & \vdots
\end{eqnarray}
and evolving for some appropriate times $t''_1$, $t''_2$, \dots, achieves the approximate evolution of the operator $e^{-i\frac{\pi}{4}Z\otimes Z}$ on any nearest-neighbour pair,
\begin{eqnarray}
U_{H_2}(t''_1) &\approx& e^{-i\frac{\pi}{4}Z\otimes Z} \otimes I \otimes I \otimes I \ldots, \label{eqn:CII}\\
U_{H_2}(t''_2) &\approx& I \otimes e^{-i\frac{\pi}{4}Z\otimes Z} \otimes I \otimes I \ldots \label{eqn:ICI} \\
 \vdots & & \vdots
\end{eqnarray}
The operator $e^{-i\frac{\pi}{4}Z\otimes Z}$ when combined with the $H$ and $T$ gates yields the controlled-not gate:
\[
\mathrm{\textsc{cnot}} = (I\otimes H) \ e^{i \pi \frac{Z\otimes Z}{4} }
(T^2\otimes T^2)
\ (I\otimes H).
\]

We have therefore shown it is possible to perform a \emph{universal} gate set using only the
Hamiltonians $H_1$ and $H_2$, although it is not clear whether this universality can be exploited in a time-efficient way.

\section{A Simple Method for Choosing Coefficients} \label{sec:Method}

The previous section argues the existence of evolution times $t_1$, $t_2$, $t_3$, \dots, for $H_1$ and $H_2$ that yield good approximations to gates in a universal set. However no constructive method was given for predicting the values of these times. In this section we give an example of a method for choosing the coefficients $a_m$, $b_m$ and $c_m$ in such a way that the appropriate evolution times have a simple, predicable form.

The coefficients $a_m$ are chosen to to have particular binary expansions, such as follows:
\begin{eqnarray}
a_1 &=& 1.00000000 \ldots \nonumber \\
a_2 &=& 0.00010000 \ldots \nonumber \\
a_3 &=& 0.00000001 \ldots \nonumber \\
\vdots && \vdots
\end{eqnarray}
That is, each successive $a_j$ is some power-of-two factor (in this case $2^4=16$) smaller.
After $H_1$ has evolved for $t=\pi/2$, the result will be a $H$ gate applied to the first qubit, and approximately the identity applied to every other qubit.
After $t=8\pi$, the
identity will have been applied to the first qubit exactly, since $e^{\pm i8\pi}=1$. The second qubit will have had an $H$ gate applied exactly, and approximately the identity will have been applied to the remaining qubits. Similarly, after $t=128\pi$, the first and second qubits
will have performed the identity exactly. The third qubit will have
undergone a $H$ gate exactly. Likewise, in general an evolution time $t=16^m\pi/2$ will yield the $H$ gate applied to qubit $m$. Parameters $b_m$ and $c_m$ can be chosen in a similar way. Higher accuracy will be yielded by replacing the value 16 throughout by some larger power of two.

The method gives us a constructive way of choosing the coefficients and switching times of $H_1$ and $H_2$ in a way that allows us to achieve a useful gate set. However, it is manifestly inefficient, requiring a time that scales exponentially with number of qubits, and requires the ability to engineer the strengths of each term to an exponential accuracy. It is open question if it is possible to find a
constructive method of choosing the coefficients and switching times of $H_1$ and $H_2$ that yields an efficient protocol. 

Even an inefficient version of the scheme might be worth implementing in certain circumstances. One could imagine a quantum computer which is divided into blocks each containing a fixed number of qubits $n$, where nearest blocks overlap by one qubit.  The universal dynamics on each block could be achieved using the scheme described above. If the block size remains fixed as the size of the quantum computer scales up, then the complexity of the scheme will also scale efficiently. This would be worth doing in circumstances where the disadvantages of increased time-complexity of applying the scheme to each $n$-qubit block are outweighed by the benifits of having just two control degrees of freedom per block.

\section{Open Problem and Conclusion} \label{sec:Conclusion}


We have identified a simple set of controls that generate universal quantum dynamics. Our controls consist of just two physically
realistic 2-local Hamiltonians. We showed that these two Hamiltonians can implement a
universal gate set, and we gave a simple (although inefficient) method for choosing
coefficients and switching times. It remains an open question whether it is possible to utilize such a simple set of universal controls it a way that is efficient and noise-tolerant. 

\section*{Acknowledgments}

The authors would like to thank Michael Nielsen whose original
question inspired this paper, and Michael Bremner for providing helpful comments on the manuscript. C.D.H.~was supported by UK EPSRC
grant number EP/C012674/1.

\bibliography{bibliography}

\end{document}